\documentclass{iopart}
\usepackage{iopams}
\usepackage{amssymb,amstext,amsfonts}
\usepackage{graphicx}

\usepackage{color}

\newcommand{\rem}[1]{}
\usepackage{bm}

\def\beq{\begin{equation}}
\def\eeq{\end{equation}}
\def\bef{\begin{figure}}
\def\eef{\end{figure}}

\begin{document}

\title[Computer simulation of finite-size broadening of solid-liquid interfaces]
{Computer simulation studies of finite-size broadening of solid-liquid
interfaces: From hard spheres to nickel}
\author{T Zykova-Timan$^{1}$, R E Rozas$^{2}$, J Horbach$^{2}$, and 
K Binder$^{1}$}

\address{$^{1}$Institut f\"ur Physik, Johannes-Gutenberg-Universit\"at Mainz,
               Staudinger Weg 7, 55099 Mainz, Germany\\
         $^{2}$Institut f\"ur Materialphysik im Weltraum, Deutsches Zentrum f\"ur
               Luft- und Raumfahrt (DLR), 51170 K\"oln, Germany}
\date{\today}

\begin{abstract}
Using Molecular Dynamics (MD) and Monte Carlo (MC) simulations
interfacial properties of crystal-fluid interfaces are
investigated for the hard sphere system and the one-component metallic
system Ni (the latter modeled by a potential of the embedded atom
type). Different local order parameters are considered to obtain order
parameter profiles for systems where the crystal phase is in coexistence
with the fluid phase, separated by interfaces with (100) orientation of
the crystal.  From these profiles, the mean-squared interfacial width
$w^2$ is extracted as a function of system size. We rationalize the
prediction of capillary wave theory that $w^2$ diverges logarithmically
with the lateral size of the system.  We show that one can estimate the
interfacial stiffness $\tilde{\gamma}$ from the interfacial broadening,
obtaining $\tilde{\gamma}\approx 0.5\, k_B T/\sigma^2$ for hard spheres
and $\tilde{\gamma}\approx 0.18$\,J/m$^2$ for Ni.
\end{abstract}
\section{Introduction}
A key issue towards the microscopic understanding of crystallization
from the melt is the knowledge about the properties of the equilibrium
crystal-melt interface. Although experimental techniques such as electron
microscopy and X-ray scattering give insight into the structure of
solid-liquid interfaces \cite{kaplan06}, at least for atomistic
systems, interfacial properties such as interfacial tensions or kinetic
growth coefficients are hardly accessible in experiments. In principle,
the situation is different for colloidal systems where microscopy allows
for the direct measurement of particle trajectories and thus, similar
as in a computer simulation, any quantity of interest can be computed
from the positions of the particles.  Recently, several experimental
studies \cite{gasser01,dullens06,prasad07,guzman09} were devoted to the
study of solid-liquid interfaces in colloidal suspensions using confocal
microscopy. However, a direct measurement of the anisotropic interfacial
tension for a solid-liquid interface has not been realized so far.
Moreover, it is an open question to what extent typical colloidal systems
such as hard spheres can serve as model systems for crystallization
processes on the atomistic scale, as they occur, e.g., in metallic alloys.

Interfacial properties are also central parameters in the continuum
modeling of crystal growth; e.g.~the widely used phase field method
needs the anisotropic interfacial tension as input (for a recent review
of the phase field approach see Ref.~\cite{emmerich08}).  The fact
that interfacial tensions are in general not known from experiments
reduces the predictive power of the phase field method.  Recently,
more microscopic approaches for the description of crystallization
processes have been proposed. The phase field crystal (PFC) method
\cite{elder02} is a generalization of phase field modeling to the
atomistic scale.  As shown by van Teeffelen et al.~\cite{teeffelen09},
the PFC method can be derived from dynamic density functional theory
(DDFT) \cite{archer04,teeffelen08}, the latter providing an ``ab initio
approach'' to dynamic crystallization and freezing phenomena. Thus,
there is hope that both PFC and DDFT will lead to some progress towards
a microscopic understanding of crystallization phenomena. We note also
that in the framework of static density functional theory, thermodynamic
properties of the crystal-melt interface such as the interfacial tension
can be predicted, at least for simple model systems (e.g.~hard spheres)
\cite{dft}.

For all the latter models, one has to keep in mind that they introduce
various severe approximations: in particular, since they are mean-field
theories, statistical fluctuations are neglected. Thus, capillary wave
excitations \cite{buff65} are not taken into account that strongly
affect the interfacial properties (e.g.~a broadening of the mean-squared
interfacial width).

Beyond mean-field theories, particle-based computer simulations are
an appropriate tool to study solid-liquid interfaces at a microscopic
level. In principle, molecular dynamics (MD) as well as Monte Carlo (MC)
simulations provide a numerically exact treatment of the statistical
mechanics, only based on a model potential that describes the interactions
between the particles. However, when one examines the details of
the solid-liquid interface at phase coexistence via such simulation
techniques, one must be aware of various problems: a slight deviation from
the correct coexistence condition, details of the averaging procedure,
insufficient sampling of statistical fluctuations, or an inappropriate
choice of the order parameter may cause more or less drastic deviations
from the correct result.  These problems have hampered progress in the
area.  On the one hand, new and presumably rather accurate methods for the
analysis of solid-liquid interfaces have been presented for hard spheres
\cite{davidchack98,davidchack00}, Lennard-Jones systems \cite{huitema99},
models of metallic systems \cite{jesson00,hoyt01,asta02,kerrache08}, and
silicon \cite{buta08}.  On the other hand, in none of these studies it
has been systematically analyzed how finite size effects influence the
properties of solid-liquid interfaces.  In the present work, we do the
first steps to fill this gap and we demonstrate that, in the framework
of capillary wave theory, the interfacial stiffness can be estimated
from an analysis of finite size effects.

A comparative study is performed of the structure of solid-liquid
interfaces of hard spheres and a model of Ni.  In this manner, we
shed light on differences between both kinds of systems and thus, we
contribute to the question whether a typical colloidal system (hard
spheres) can serve as a model for a typical metallic system (Ni) with
respect to the interfacial properties at coexistence. Furthermore, by
using both MC and MD simulations we rationalize that both methods can
be applied to the interfacial analysis.

The outline of the paper is as follows. In the next section, we
introduce the models used for the simulations and summarize the
simulation methodology. In Sec.~3 we give a brief overview over the
results of capillary wave theory that we use to determine the interfacial
stiffness. Then, in Sec.~4 we present the results for the fluid structure,
the local order parameter profiles and the system size dependence of the
mean-squared interfacial width from which we estimate the interfacial
stiffness.  Finally, we summarize the results.

\section{Models and simulation techniques}
In this section, we introduce briefly the main details of the
MC simulations for the hard sphere system and the MD simulations
for Ni. Moreover, for both cases we describe how we have prepared
configurations in a slab geometry where the crystal phase in the middle of
an elongated (rectangular) simulation box is at coexistence with the fluid
phase, separated by two interfaces (parallel to the $xy$-plane)
and using periodic boundary conditions in all three spatial directions.

\subsection{Monte Carlo simulation of hard spheres}
The hard sphere system is defined via the potential
\begin{equation}
\label{eq_hsmodel}
u(r) = \left\{
\begin{array}{ll}
\infty & r<\sigma \\
0      & r\geq\sigma \, ,
\end{array}
\right.
\end{equation}
with $r$ the distance between two particles and $\sigma$ the diameter of
a particle. Throughout the paper all length scales are measured in 
units of $\sigma$ for the hard sphere system.

Since for any allowed hard sphere configuration, the total potential
energy is zero, temperature $T$ is only a scaling factor and the
thermodynamic properties are fully controlled by the packing density
$\eta=\frac{\pi\sigma^3}{6}\frac{N}{V}$ (or, when the total volume $V$
of the system is a fluctuating variable, by the pressure $P$).  As a
consequence, the phase behavior of hard sphere systems is completely
driven by entropy \cite{frenkel99}. As first claimed by Kirkwood
\cite{kirkwood51}, hard spheres exhibit a fluid-to-solid transition.
However, Kirkwood's prediction was based on misleading arguments and so
it was a surprising discovery when the freezing of hard spheres was first
observed in early computer simulations \cite{alder57,wood57}. Since the
phase behavior of hard spheres depends only on packing density $\eta$,
the phase diagram is particularly simple. Due to the absence of attractive
interactions, the hard sphere model (\ref{eq_hsmodel}) does not exhibit
a liquid-vapor transition. Fluid and fcc crystal coexist between the
freezing point $\eta_{\rm f}=0.494$ and the melting point $\eta_{\rm
m}=0.545$, while the pure crystal is the stable phase for $\eta >
\eta_{\rm m}$ \cite{hoover68,frenkel-book}.

The MC simulations were carried out in the isothermal-isobaric ($NPT$)
and ($NP_zT$) ensemble, in which the pressure $P$, the temperature $T$
and the number of particles $N$ are constant (in the $NP_zT$ ensemble,
$P_z$ is the diagonal component of the pressure tensor perpendicular to
the $xy$ plane). A MC code was developed applying the standard Metropolis
algorithm \cite{binder-book,frenkel-book,krauth-book}.  The trial moves
are particle displacements, where each particle was attempted to be
displaced once per MC cycle, and system's volume rescaling that was
executed once per MC cycle. The maximum displacement is chosen to keep
the acceptance rate at 30\% for the particles and 10\% for the volume.

In test runs, we have reproduced the solid and liquid branches of
the equation of state and have found full agreement with analytical
results \cite{carn69,hall72}. Additionally, the radial distribution
function and the static structure factor for the bulk liquid has
been compared to the Percus-Yevick approximation \cite{hansen86}.
The coexistence pressure was found by the interface velocity
analysis similar to Ref.~\cite{kerrache08}. At coexistence the total
volume of the melt-crystal system is constant, since the melt is in
equilibrium with the crystal, therefore the interface velocity is
$v_{\rm I}=0$. Hence, a series of runs were performed in a wide range
of pressures.  At each simulation, the interface velocity $v_{\rm I}$
was estimated from the slope $dV/dt$ of the temporal change of the
system volume $V(t)$ in the stationary growth interval. Following
this procedure, first, we have estimated the rough location of the
coexistence pressure and then increasing the resolution of the pressure
interval we improved the initial accuracy.  The final result yielded
$p_c=11.55\pm 0.04\; k_BT/\sigma^3$, very close to the literature
value $p_c=11.567\; k_BT/\sigma^3$\cite{frenkel-book}. Similar values
($p_c=11.54, 11.5, 11.53, 11.55\; k_BT/\sigma^3$) have been reported in
Ref.~\cite{davidchack98,noya08,bruce00,dij06} accordingly.  The average
densities were found to be $\rho_{\rm s}=1.04$ for the solid and
$\rho_{\rm l}=0.9385$ for the liquid, close to the results of Hoover
and Ree \cite{hoover68} ($\rho_{\rm s}=1.04$ for solid and $\rho_{\rm
l}=0.939$). The bulk interplanar distance between lattice planes in
the crystalline phase is $\sim$ 0.784\,$\sigma$. Further details are
discussed elsewhere \cite{tanya09}.

To generate crystal-melt interfaces at coexistence, we first prepared
solid-liquid ``sandwiches'' where the (100) direction of the fcc phase is
oriented perpendicular to the $z$ axis. In this study, we considered the
system sizes of side length $L=na$ lattice spacings with $n=5,6,7,8,10$
(where $a=1.567\,\sigma$ at coexistence). The total number of particles is
$N= 2500$, 4320, 6860, 10240, 14580, and 20000, respectively.  We have
verified that the elongation $L_{z}=5L$ along the $z$ direction is
sufficient to avoid interactions between the interfaces due to periodic
boundary conditions.  For smaller elongations in $z$ direction we see
significant finite size effects in density profiles due to the interaction
of the two interfaces via periodic boundary conditions.

As a first step, a solid slab of size ($L \times L \times 3L$)
and a liquid box of size ($L \times L \times 2L$) were equilibrated
separately for 10$^{6}$ MC cycles in the $NPT$ ensemble at $p_c$.
Here, $L$ corresponds to the bulk solid density $\rho_{\rm s}$ at
coexistence. For the liquid, the same values of $L$ were used in $x$
and $y$ direction. Then, the correct bulk density of the liquid was
achieved by using simulations in the $NP_zT$ ensemble.  Next, the two
parts were placed together in a simulation box of size $L_{x}\times
L_{y}\times L_{z}$ with $L_{x}=L_{y}=L$.

The most delicate point of the preparation is how to match the solid
and fluid parts.  The initial configuration might be refused if the
melt and the crystal are too close, otherwise a large gap between
them would cause lower fluid density and artifacts with respect to
interface properties. Therefore, new fluid-solid configurations were
relaxed in $NP_zT$ simulations until the coexistence densities were
recovered in the bulk regions. The first $10^5$ MC cycles the lateral
sizes and the positions of the solid particles were fixed in order to
avoid internal stresses in the solid.  Furthermore, for each system size
we have performed isochoric runs of $2\cdot 10^5$ MC cycles, initially
rescaling the length of the box $L_z$ to the average value $\langle L_z
\rangle$, as computed in the last $5\cdot 10^4$ MC cycles of previous
$NP_zT$ runs.  Next, we have selected 50 independent configurations
for sizes $n=5,6,7,8$ and 10 independent configurations for $n=10$ to
compute the equilibrium properties of the interfaces over $5\cdot 10^4$
MC cycles in the isochoric ensemble. The statistics was collected every
20 MC cycles.  The length of the final runs was chosen to prevent the
diffusive motion of the interface, however in several configurations an
additional interface broadening was detected due to the displacement of
the solid-liquid interface. Those runs were replaced by additional runs.

\subsection{Molecular Dynamics simulation of Ni}
For Ni, a similar methodology as for the hard sphere system has been
employed. But MD simulations instead of MC simulations were performed.
As a potential to model the interactions between the particles in Ni
we used a potential of the embedded atom type, as proposed by Foiles
\cite{foiles85}. We show elsewhere \cite{rozas09} that this model
potential reproduces very well various thermodynamic and transport
properties, as obtained from experiments of liquid Ni.

As before for the hard sphere system, an inhomogeneous solid-liquid
system is simulated in a sandwich geometry with $L\times L \times L_z$
being the size of the simulation box. Again, $L_z=5L$ is chosen.  In the
following, $L$ will be also expressed in terms of the number of lattice
planes of the fcc crystal, $n$, as $L=na$ (note that the lattice constant
is given by $a=3.58$\,\AA~at the melting temperature $T_{\rm m}$).

Newton's equations of motion are integrated with the velocity form
of the Verlet algorithm \cite{book-allen}, using a time step of 1\,fs.
The melting temperature of the model is again estimated from an interface
velocity analysis, considering an inhomogeneous solid-liquid system in
the $NPT$ ensemble.  The melting temperature is obtained from the linear
fit of the interface velocity versus temperature up to an undercooling
of about $40$\,K.  From these simulations \cite{rozas09}, we have found
the melting temperature $T_{\rm m}=1748$\,K, which is in good agreement
with the experimental value, $T_{\rm m}=1726$\,K \cite{massalski86}.
Simulations of different system sizes indicate that finite size effects
are weak \cite{rozas09}, as far as the determination of $T_{\rm m}$ is
concerned. For the Ni model used in this work, the melting temperature
of the smallest system with $N = 2500$ particles was $0.5 \%$ higher than
the estimated melting temperature in the thermodynamic limit, $T_{\rm m}
= 1748$\,K.

To prepare an inhomogeneous system with two crystal-liquid interfaces
at $T_{\rm m}$, the following steps are involved: First, atoms,
disposed on a fcc lattice, are relaxed in a $NPT$ simulation for
about 30\,ps at $T_{\rm m}$ and zero pressure.  Temperature was kept
constant by coupling the system to a stochastic heat bath, i.e.~by
reassigning every 200 steps new velocities to each particle according
to a Maxwell-Boltzmann distribution. To keep the pressure constant,
an algorithm proposed by Andersen was used, setting the mass of the
piston to 100\,eV\,ps$^2$/\AA$^2$ \cite{andersen80}.  In the next step,
a liquid and a crystal region are defined in the system such that the
crystal region in the middle of the elongated simulation box occupies a
volume of $L^3$.  Atoms in the crystal region remain at fixed positions
while the rest of the system is heated up to 2400\,K which is well above
the melting point. At this step, only volume changes with respect to the
expansion and compression of the box in the direction perpendicular to
the liquid-solid interface are applied, i.e.~the simulation is done in
the $NP_zT$ ensemble.  In order to completely melt the liquid regions,
the simulation runs over about 100\,ps.  Then, the temperature of the
melt is set back to the initial temperature at which the crystal was
prepared. A run over 50\,ps in the $NP_zT$ ensemble is added where all
the particles are allowed to move. From the last 10\,ps of this run
the average length of the box in $z$ direction, $\langle L_z \rangle$,
is determined. After the length of the simulation box in $z$ direction
is rescaled to $\langle L_z \rangle$, the simulation continues with a
run over 20\,ps in the $NVT$ ensemble.  From the last 10\,ps of this
run, the average total energy of the system is computed. The system
is set to this energy by rescaling the velocities of the particles
appropriately. Finally, microcanonical production runs over 1\,ns are
done from which the information about the interfacial properties are
obtained. We considered systems of lateral size $L=na$ with $n=5$, 6,
7, 8, 9, 10, 11, 12, 13, 14, and 15. The total number of particles in
these systems is $N= 2500$, 4320, 6860, 10240, 14580, 20000, 26620,
34560, 43940, 54880, and 67500, respectively. For each system size, 
5 independent runs were performed.

\section{Capillary fluctuations}
\label{tanh}

In this work, we consider systems where a crystal phase is at coexistence
with a fluid phase and the two phases are separated from each other
by interfaces (two interfaces are formed due to periodic boundary
conditions). As a matter of fact, the presence of the interfaces
breaks the translational invariance, which is a continuous symmetry
property of the underlying Hamiltonian. The latter leads to the occurrence
of Goldstone excitations: long-wavelength transverse excitations,
known as capillary waves, appear that are thermally driven
undulations of the interface. At infinite wavelength (i.e.~in the limit
of wavenumber $q\to 0$), they describe an overall translational motion
of the interface with zero energy cost. In the framework of mean-field
approaches such as phase-field modeling, capillary wave excitations are
neglected, although they strongly affect the interfacial properties.

This can be seen in the framework of capillary wave theory (CWT)
\cite{buff65}. This theory describes the free energy cost $\Delta F$
of long-wavelength undulations of an interface. For three-dimensional
systems, CWT predicts a logarithmic divergence of the mean squared
width, $w^2$, of the interface with the lateral system size
$L$. As we see below, in a computer simulation, this result of
CWT can be used as a method to determine the interfacial tension
by measuring the mean-squared width $w^2$ for different system
sizes. Whereas this method has been successfully applied to the
Ising model \cite{schmid92,hasenbusch92,mueller05}, polymer mixtures
\cite{werner97,werner99a,werner99b}, and liquid-vapor interfaces in the
Asakura-Oosawa model for colloid-polymer mixtures \cite{vink04,vink05},
it has not been used as a method to estimate the interfacial tension
of crystal-fluid interfaces. In the latter case, many of the recent
simulation studies on hard spheres (e.g.~\cite{davidchack98}), and
metallic systems (e.g.~\cite{hoyt01}) have used an analysis of the
capillary wave spectrum to determine the interfacial tension. However,
here, we demonstrate that also in the case of solid-liquid interfaces the
interfacial stiffness can be computed by measuring $w^2$ as a function
of $\ln L$. Both for polymer mixtures \cite{werner97,werner99a} and
the liquid-vapor transition of the Asakura-Oosawa model \cite{vink05},
it has been shown that the analysis of the capillary wave spectrum and
the finite-size analysis of the interfacial broadening yield values for
the interfacial tension that are in agreement.

As before, we consider atomically rough crystal-fluid interfaces. We
parametrize the local fluctuations of the interface by a function $h(x,y)$
that denotes the local deviation of the interface position $z_0(x,y)$
from the mean value $h(x,y)=z_0(x,y)-\langle z_0(x,y) \rangle$ (here,
$z$ is the Cartesian component perpendicular to the interface, $x$, $y$
the ones parallel to it). $h(\vec{\rho})\equiv h(x,y)$ can be expressed
in Fourier coordinates, $h(\vec{\rho})= \sum_{\vec{q}} h(\vec{q}) \exp(i
\vec{q}\cdot\vec{\rho})$ (with the wavevector $\vec{q}=(q_x,q_y)$). Then,
the total free energy of the interface can be expressed in reciprocal
space as
\begin{equation}
  \label{eq_deltaf}
  \Delta F = \frac{L^2\tilde{\gamma}}{2} \sum_{\vec{q}} q^2
   |h_{\vec{q}}|^2 \, .
\end{equation}
Here, $L^2$ is the area of the flat interface. $\tilde{\gamma}$ is
the interfacial stiffness, defined by $\tilde{\gamma} = \gamma + d^2
\gamma/ d\theta^2$ with $\gamma$ the interfacial tension and $\theta$
the angle between the interface normal and the (100) direction. The
interfacial stiffness takes into account the anisotropy of the
interfacial tension in case of a crystal-fluid interface; of course,
in case of a liquid-vapor interface $\tilde{\gamma}$ would be replaced
by $\gamma$ in Eq.~(\ref{eq_deltaf}).

Since the different $q$ modes in Eq.~(\ref{eq_deltaf}) are decoupled, it
follows from the equipartition theorem that each mode carries an energy of $k_BT$
and thus one obtains:
\begin{equation}
\label{eq_ampl}
\langle | h_{\vec{q}}|^2\rangle = \frac{k_B T}{L^2 \tilde{\gamma} q^2} \, .
\end{equation}
This expression can be used to determine $\tilde{\gamma}$ by measuring the
slope of the straight line that fits $1/\langle | h_{\vec{q}}|^2\rangle$
plotted as a function of $q^2$. In fact, this has been done in recent
simulation studies of hard spheres \cite{laird06} and metallic 
systems \cite{hoyt01}. However, in
the latter works, a geometry with $L_y<<L_x$ were chosen such that only
the fluctuations of a quasi one-dimensional ``ribbonlike'' interface
are considered.  Although this simplifies the analysis it also alters
the nature of the capillary fluctuations (see below). In this work,
we consider therefore only a geometry with $L=L_x=L_y$.

Equation (\ref{eq_ampl}) can be used to determine the mean-squared
interfacial width $w^2$, given by
\begin{equation}
w^2_{\rm cw} = \langle | h_{\vec{\rho}}|^2\rangle =
\sum_{\vec{q}} \langle | h_{\vec{q}}|^2\rangle
= \frac{L^2}{(2\pi)^2} \int d\vec{q} \; \langle | h_{\vec{q}}|^2\rangle \; ,
\end{equation}
which yields
\begin{equation}
 \label{eq_fe}
w^2_{\rm cw} = \frac{k_B T}{2\pi \tilde{\gamma}}
\int_{2\pi/L}^{2\pi/\ell} \frac{dq}{q}
= \frac{k_B T}{2\pi \tilde{\gamma}} \ln (L/\ell) \; .
\end{equation}
In Eq.~(\ref{eq_fe}), $\ell$ is a cut-off length that is introduced in
accordance with the assumption that only modes with a wavelength larger
than the typical width of the interface are taken into account.  Note that
with the aforementioned quasi one-dimensional ribbonlike interface the
mean-squared width would increase linearly with $L$.

In mean-field theory the interface between coexisting phases is assumed
to be flat and the interfacial profile $\phi(z)$ is described by a
hyperbolic tangent,
\begin{equation}
\label{eq_mf}
\phi(z) = A + B \tanh\left(\frac{z-z_0}{w_0}\right)
\end{equation}
where $A$ and $B$ are parameters related to the bulk values of the
densities or order parameters, $z_0$ and $w_0$ are the position of the
interface and its width, respectively.

Now, the idea is to combine the mean-field result with that of CWT by
considering $w_0$ as the width of an intrinsic profile that is superposed
by fluctuations described by $w^2_{\rm cw}$. The total width of the
profile is then obtained from a convolution approximation \cite{jasnow84},
\begin{equation}
\label{eq_width}
w^2 = w_0^2 + \frac{\pi}{2} w^2_{\rm cw} = w_0^2 + 
\frac{k_B T}{4\tilde{\gamma}} \ln L
- \frac{k_B T}{4\tilde{\gamma}} \ln \ell \; .
\end{equation}
When using this equation as a fit formula to analyse profiles
as measured in the simulation, it is impossible to disentangle
the intrinsic width contribution $w_0^2$ from the ``cut-off''
contribution $-\frac{k_B T}{4\tilde{\gamma}} \ln \ell$.  This issue
has been discussed in detail for the case of polymer mixtures
\cite{werner97,werner99a,werner99b,kerle99,binder00}.

The main issue in the following is to investigate whether
Eq.~(\ref{eq_width}) can be used to measure the interfacial stiffness
in a computer simulation.  To this end, fits with Eq.~(\ref{eq_mf})
to order parameter profiles are used to obtain an effective interface
width for different lateral system sizes $L$.  This will be described
in the next section.

\section{Results}
\subsection{Static structure factor: Hard sphere fluid vs.~Ni melt}
\begin{figure}
\begin{center}
\includegraphics[width=0.5\textwidth,clip]{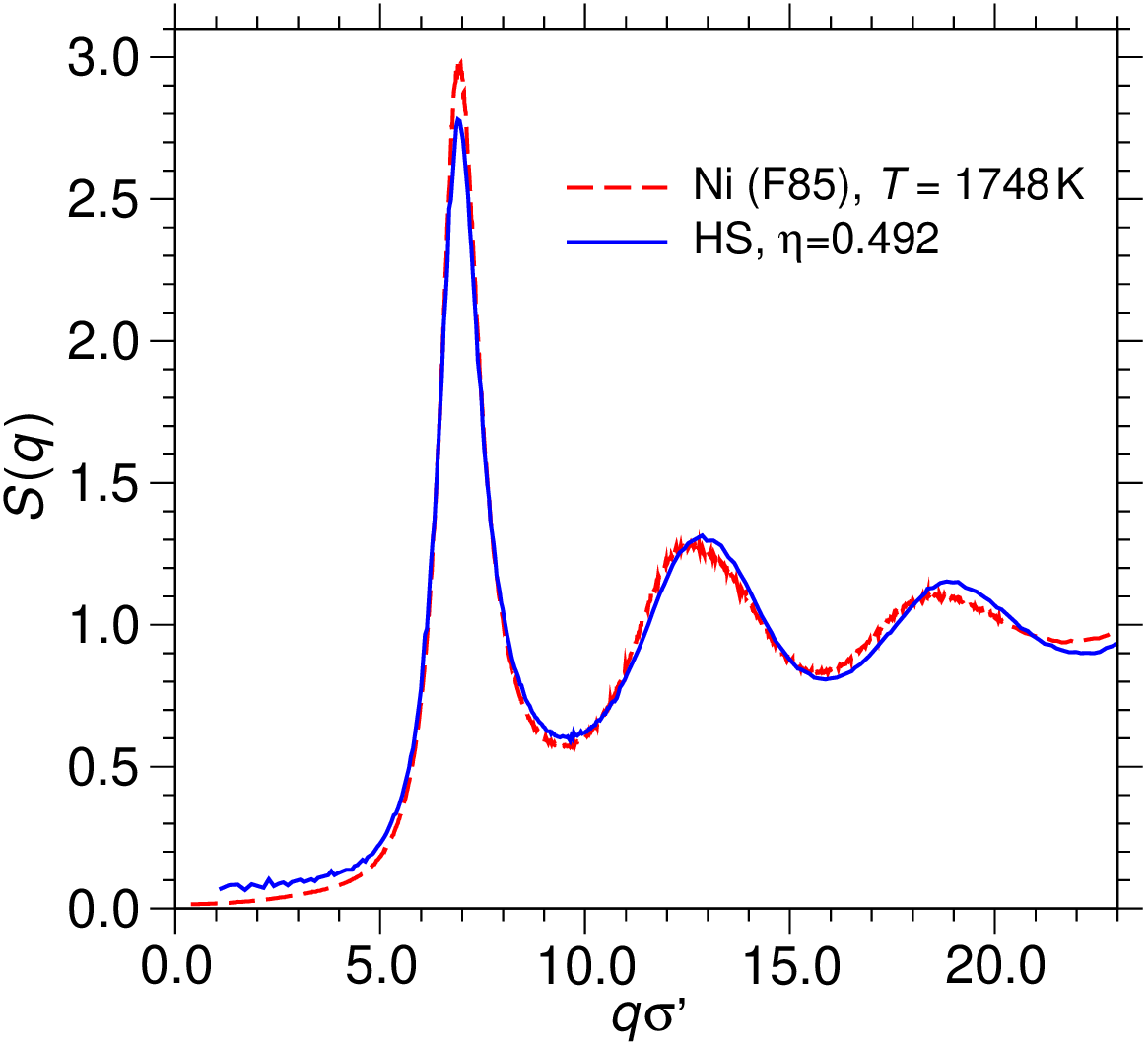}
\caption{Static structure factor $S(q)$ of the hard sphere system and the
Ni melt at coexistence, i.e.~at the volume packing fraction $\eta=0.492$
for the hard spheres and at the temperature $T=1748$\,K for Ni.}
\label{fig1}
\end{center}
\end{figure}
In order to study the structural differences between the bulk hard sphere
fluid and the bulk Ni melt at coexistence, Fig.~\ref{fig1} displays the static
structure factor \cite{hansen86} for both systems,
\begin{equation}
S(q) = \frac{1}{N} \left\langle {\Big |} \sum_{k=1}^{N} \exp[i \vec{q} 
\cdot \vec{r}_k] {\Big |}^2 \right\rangle 
\end{equation}
with $\vec{r}_k$ the position of particle $k$ and $\vec{q}$ the
three-dimensional wavevector (different from the two-dimensional
wavevectors that we have considered in the previous section). To
provide a better comparison between the structure factors
for the two systems, we have multiplied the wave-number $q$ in
Fig.~\ref{fig1} by $\sigma^\prime=\sigma$ for the hard sphere system
and $\sigma^\prime=2.24$\,\AA~for Ni (which is similar to the nearest
neighbor distance, $r_{\rm NiNi}=2.42$\,\AA~\cite{rozas09}).

Mainly two differences between the two systems can be inferred from
Fig.~\ref{fig1}.  Towards $q\to 0$ the structure factor for Ni has a much
lower amplitude, indicating that, at coexistence the Ni melt has a lower
compressibility than the hard sphere system. Moreover, the amplitude
of the first peak in $S(q)$ for Ni is slightly larger. But the overall
shape of $S(q)$ is surprisingly similar for both systems which shows that the
hard sphere system's structure is very similar to that of Ni.

\subsection{The structure of the solid-fluid interfaces}
\begin{figure}
\begin{center}
\includegraphics[width=0.45\textwidth,clip]{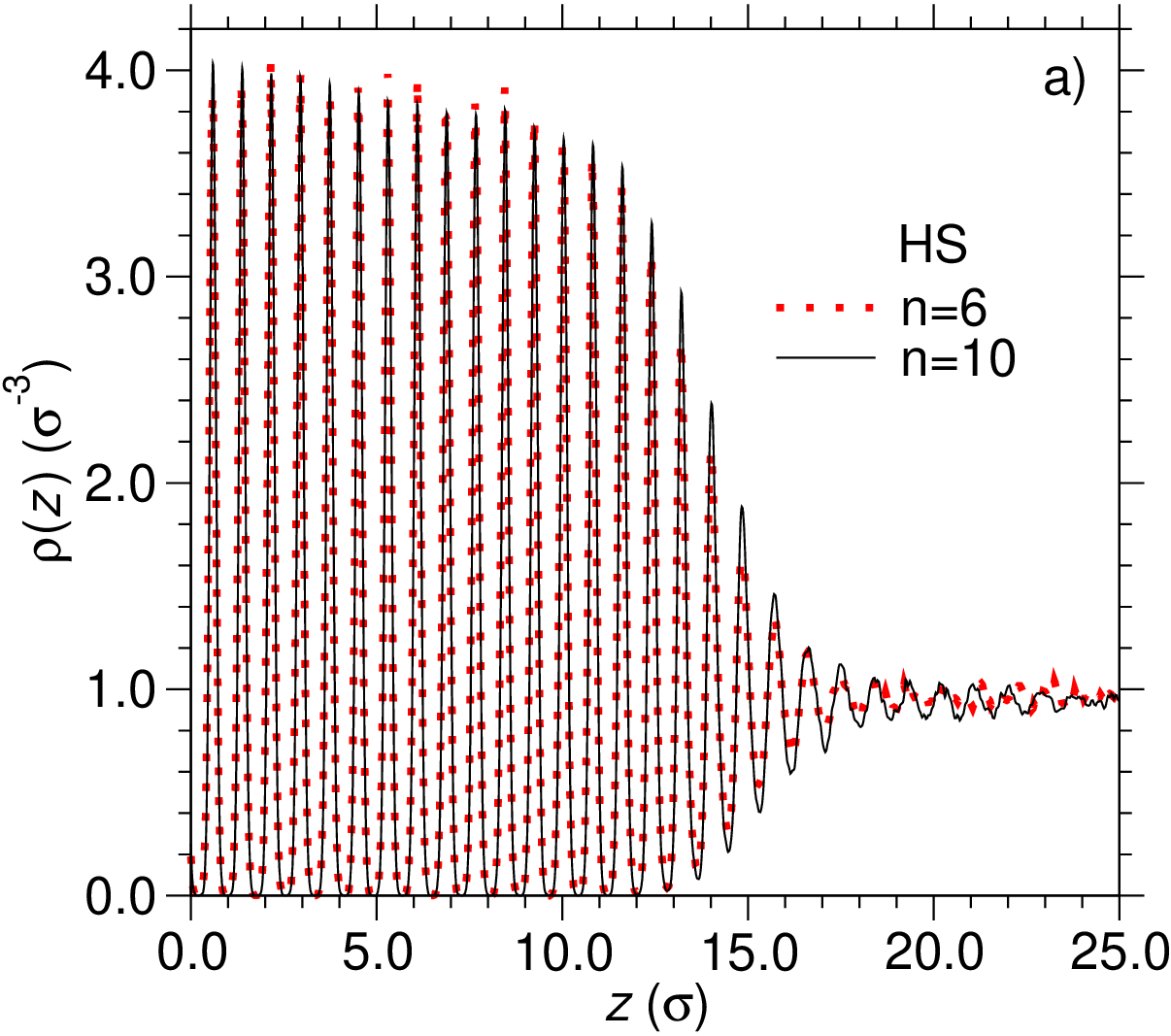}
\includegraphics[width=0.46\textwidth,clip]{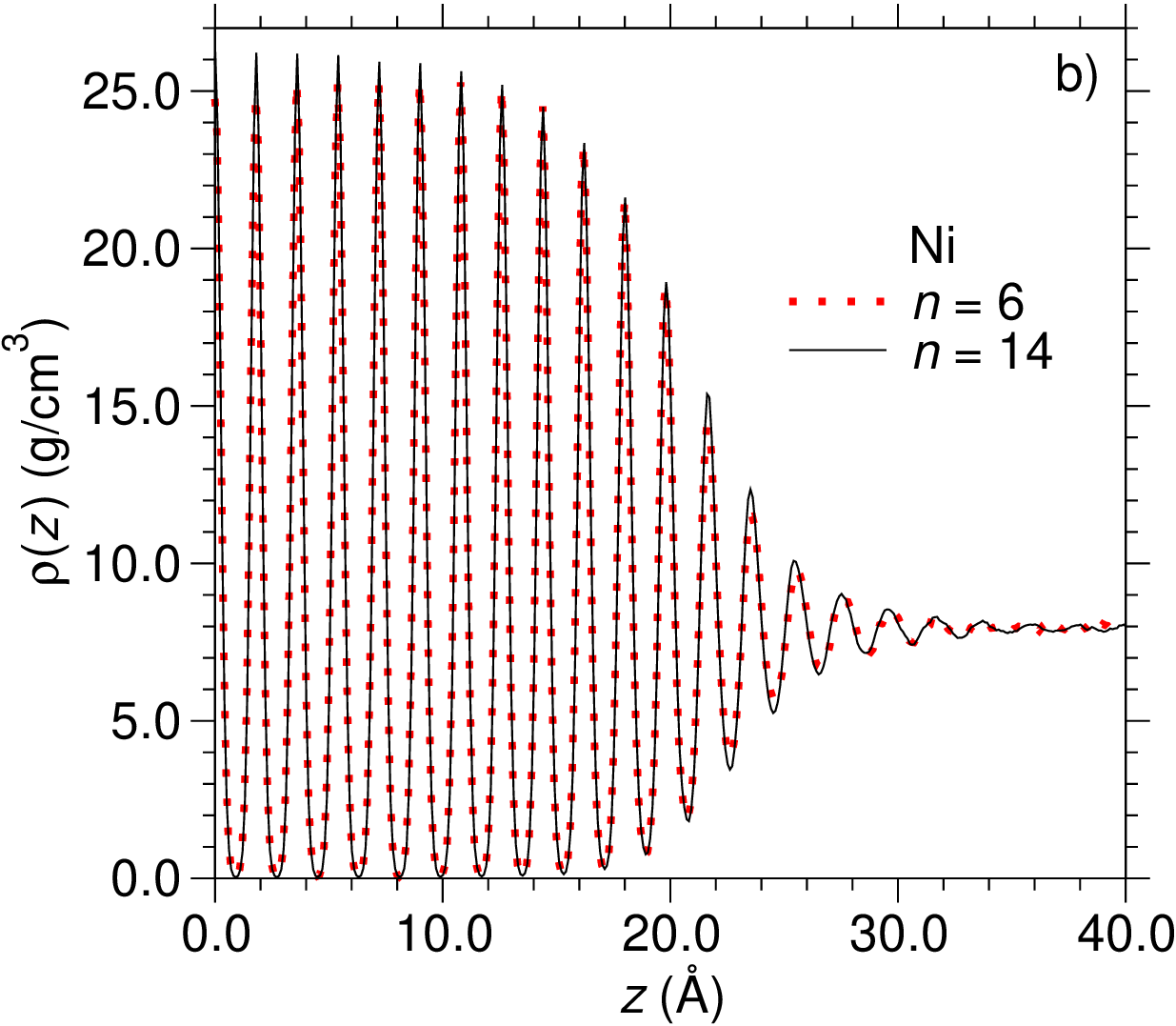}
\caption{Density profiles for a) hard spheres and b) Ni. In each
case, the profiles are shown for two different system sizes.}
\label{fig2}
\end{center}
\end{figure}
A simple quantity to characterize the structure of the interface
is provided by the density profile $\rho(z)$ across the interface.
To compute $\rho(z)$, the system is divided into slices of thickness
$\Delta z$ and then, in each slice, one counts the number of particles
and divides by the volume of the slice $L^2\Delta z$.  The displacement
of the lattice planes along the $z$-axis during the time evolution was
corrected for each configuration.  As for the order parameter profiles
shown below, we have averaged the density profiles over the two interfaces
in each system that are present due to periodic boundary conditions.

Figure~\ref{fig2} shows density profiles for the hard sphere system and
Ni, in each case for two system sizes.  The shape of the profiles is
typical for inhomogeneous systems with crystal-fluid interfaces. 
In fact, very similar density profiles have been found for various
systems with solid-liquid interfaces \cite{davidchack98,huitema99,jesson00,kerrache08,buta08}.
Whereas
one observes huge oscillations in the crystalline region due to the
presence of crystalline layers, in the fluid region $\rho(z)$ is
constant. In between, i.e.~in the interface region, the amplitude of
the peaks decreases.  Obviously, the size effects are very small for
both considered systems.  However, for the big systems the height of the
peaks is slightly larger in the interface region.  We note that although
the density profiles have a very different shape in the solid and the
liquid regions, both for the HS system and for Ni the differences in the
average densities of the solid and the liquid phase are rather small. In
the HS case, the solid density is about 10\% higher ($\rho_{\rm s}=1.04$,
$\rho_{\rm l}=0.9385$) and for Ni, it is about 5\% higher ($\rho_{\rm
s}=8.357$\,g/cm$^3$, $\rho_{\rm l}=7.928$\,g/cm$^3$).
Due to this small difference between $\rho_{\rm l}$ and $\rho_{\rm s}$
the density is not a good order parameter for the investigation 
of interfacial properties such as the interfacial width.

\begin{figure}
\begin{center}
\includegraphics[width=0.45\textwidth,clip]{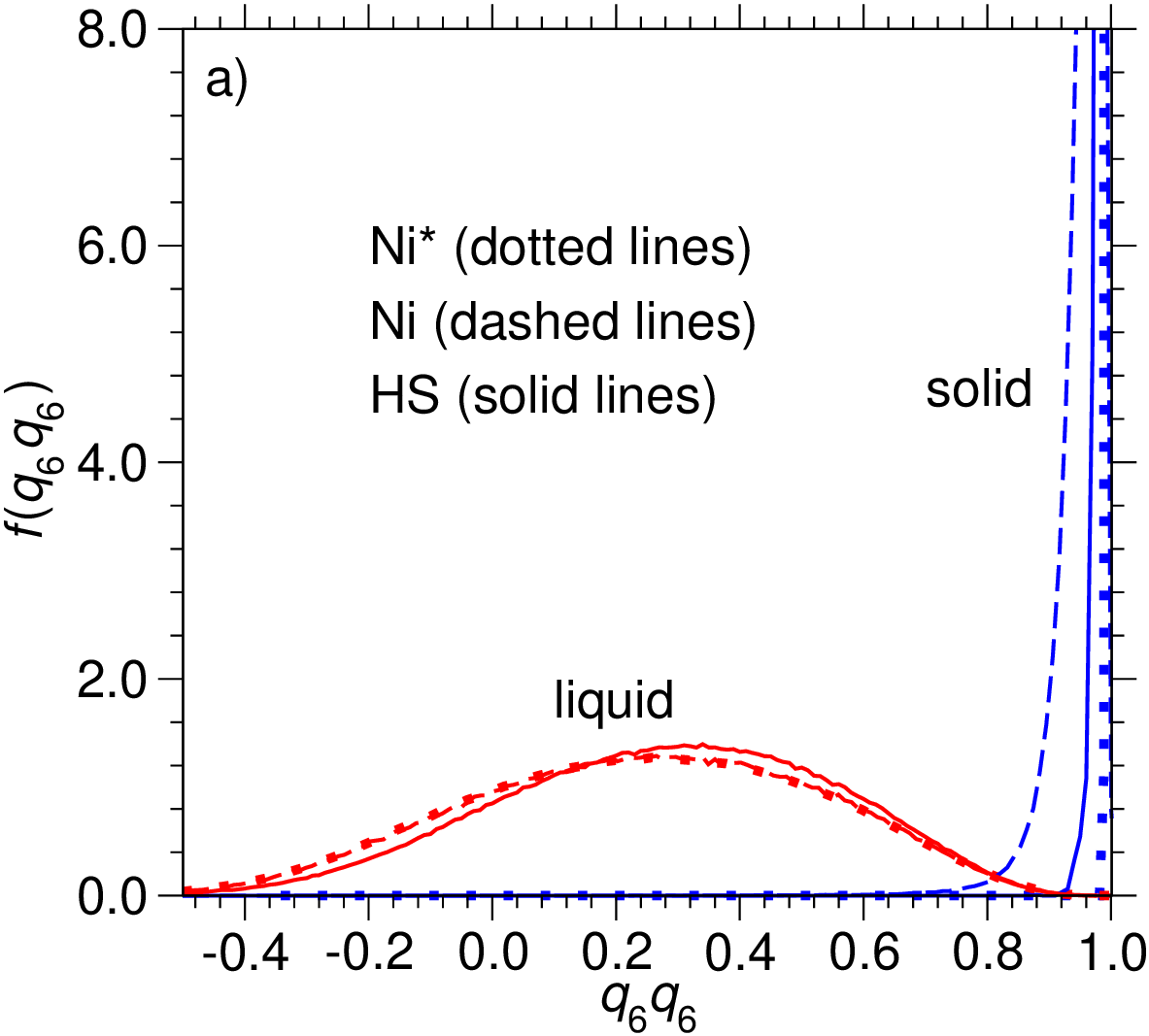}
\includegraphics[width=0.45\textwidth,clip]{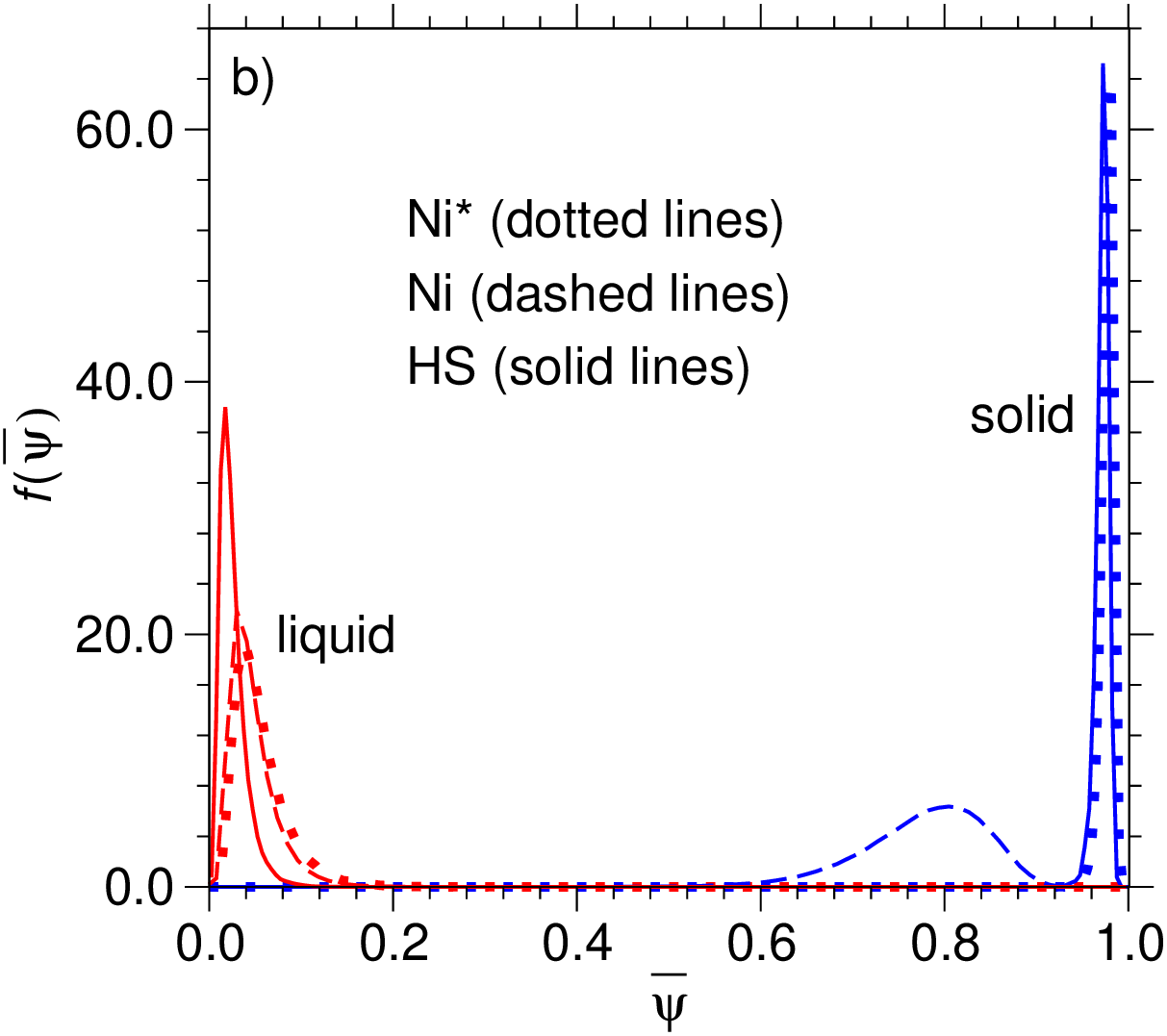}
\caption{Probability distributions of the order parameter in Ni (broken
lines) and the hard sphere system (solid lines) for the liquid and the
fcc phase, as indicated; a) $q_6q_6$, b) $\bar{\Psi}$. Different time
intervals for obtaining the average particle positions are considered
for Ni; thereby, the dashed lines correspond to an interval of 0.1\,ps
and the dotted lines to 1\;ps (the latter case is referred to as $Ni^*$).}
\label{fig3}
\end{center}
\end{figure}
Another possibility to characterize the structure of interfaces
is provided by profiles of local order parameters. Steinhardt et
al.~\cite{steinhardt83} have proposed rotational-invariant order
parameters in terms of expansions into spherical harmonics $Y_{lm}$,
\begin{equation}
\label{eq:localorder2}
Q_l(i)=\left(\frac{4\pi}{2l+1}\sum_{m=-l}^{l}{|\bar{Q}_{lm}|^2}\right)^{1/2}
\end{equation}
with
\begin{equation}
\label{eq:localorder1}
\bar{Q}_{lm}(i)=
\frac{1}{Z_i}
\sum_{j=1}^{Z_i}{Y_{lm}(\theta(\vec{r}_{ij}),\phi(\vec{r}_{ij}))} \, ,
\end{equation}
where $\vec{r}_{ij}$ is the distance vector between a pair of neighboring
particles $i$ and $j$, $Z_i$ is the number of neighbors within a given
cut-off radius, and $\theta(\vec{r}_{ij})$ and $\phi(\vec{r}_{ij})$
are the polar bond angles with respect to an arbitrary reference frame.

Similar local order parameters have been introduced
by ten Wolde et al.~\cite{tenwolde95}, defined by
\begin{equation}
\label{eq:localorder3}
q_lq_l(i)=\frac{1}{Z_i}\sum_{j=1}^{Z_i}{{\bf q}_l(i)\cdot{\bf q}_l(j)} \, .
\end{equation}
The internal product in this sum is given by
\begin{equation}
\label{eq:localorder4}
{\bf q}_l(i)\cdot{\bf q}_l(j)=\sum_{m=-l}^{l}{\tilde{q}_{lm}(i)\tilde{q}_{lm}(j)^*}
\end{equation}
with
\begin{equation}
\label{eq:localorder5}
\tilde{q}_{lm}(i)=
\frac{\bar{Q}_{lm}(i)}
{\left(\sum_{m=-l}^{l}{|\bar{Q}_{lm}(i)|^2}\right)^{1/2}} \, .
\end{equation}
In the following, we use the parameter $q_6q_6$ that is defined 
by Eqs.~(\ref{eq:localorder3})-(\ref{eq:localorder5}), setting $l=6$.

Another local order parameter used in this work was introduced by 
Morris \cite{morris02}
\begin{equation}
\label{eq:localorder6}
\Psi(i)={\Big |} \frac{1}{N_q}\frac{1}{Z_i}
\sum_{j=1}^{Z_i}\sum_{k=1}^{N_q}
\exp(i\vec{q}_k\cdot \vec{r}_{ij}) {\Big |}^2
\end{equation}
where the wavevectors $\vec{q}_k$ are chosen such that in a perfect 
crystal
\begin{equation}
\label{eq:localorder7}
|\exp(i\vec{q}_k\cdot \vec{r}_{ij})|=1 \, .
\end{equation}
Again, $\vec{r}_{ij}$ is the distance vector between neighboring particles.
With respect to the basis vectors of the fcc lattice with lattice
constant $a$, $\vec{a}_1=a/2(1,1,0)$, $\vec{a}_2=a/2(0,1,1)$
and $\vec{a}_3=a/2(1,0,1)$, an appropriate choice of wavevectors
is $\vec{b}_1=2\pi/a(-1,1,-1)$, $\vec{b}_2=2\pi/a(1,-1,1)$ and
$\vec{b}_3=2\pi/a(1,1,-1)$. An additional average of $\Psi(i)$ over a
particle with index $i$ and its neighboring particles yields
\begin{equation}
\label{eq:localorder8}
\bar{\Psi}(i)=\frac{1}{Z_i+1}\left(\Psi(i)+\sum_{j=1}^{Z_i}\Psi(j)\right) \; .
\end{equation}
The parameter $\bar{\Psi}$ together with $q_6q_6$ is used in the following
to distinguish solid particles from fluid particles and to identify the
interfacial regions. To select the nearest neighbors we introduced the 
cut-off radii that correspond to the first minimum of the radial
distribution function of the bulk liquid phase at coexistence.

%
\begin{figure}
\begin{center}
\includegraphics[width=0.45\textwidth,clip]{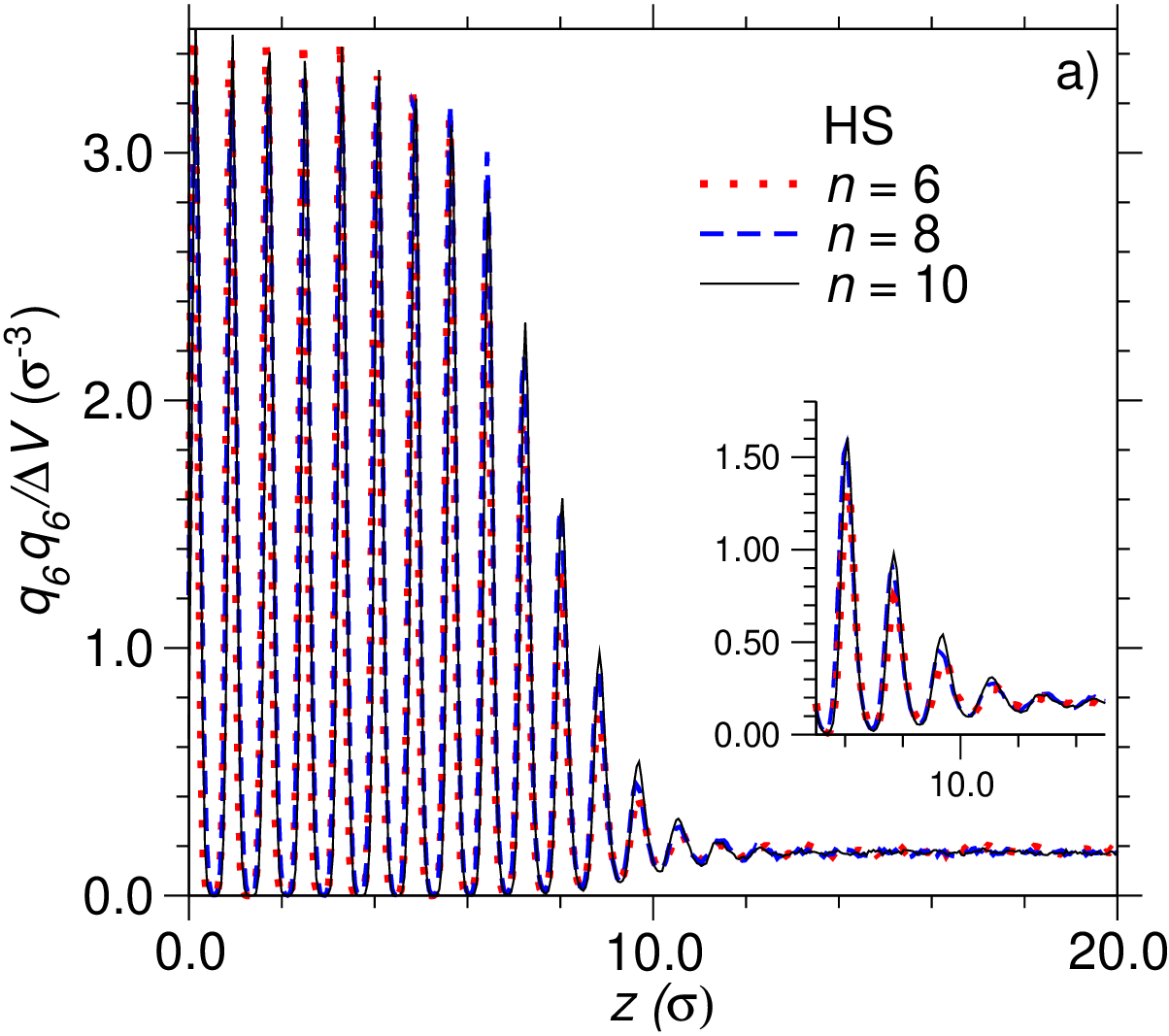}
\includegraphics[width=0.45\textwidth,clip]{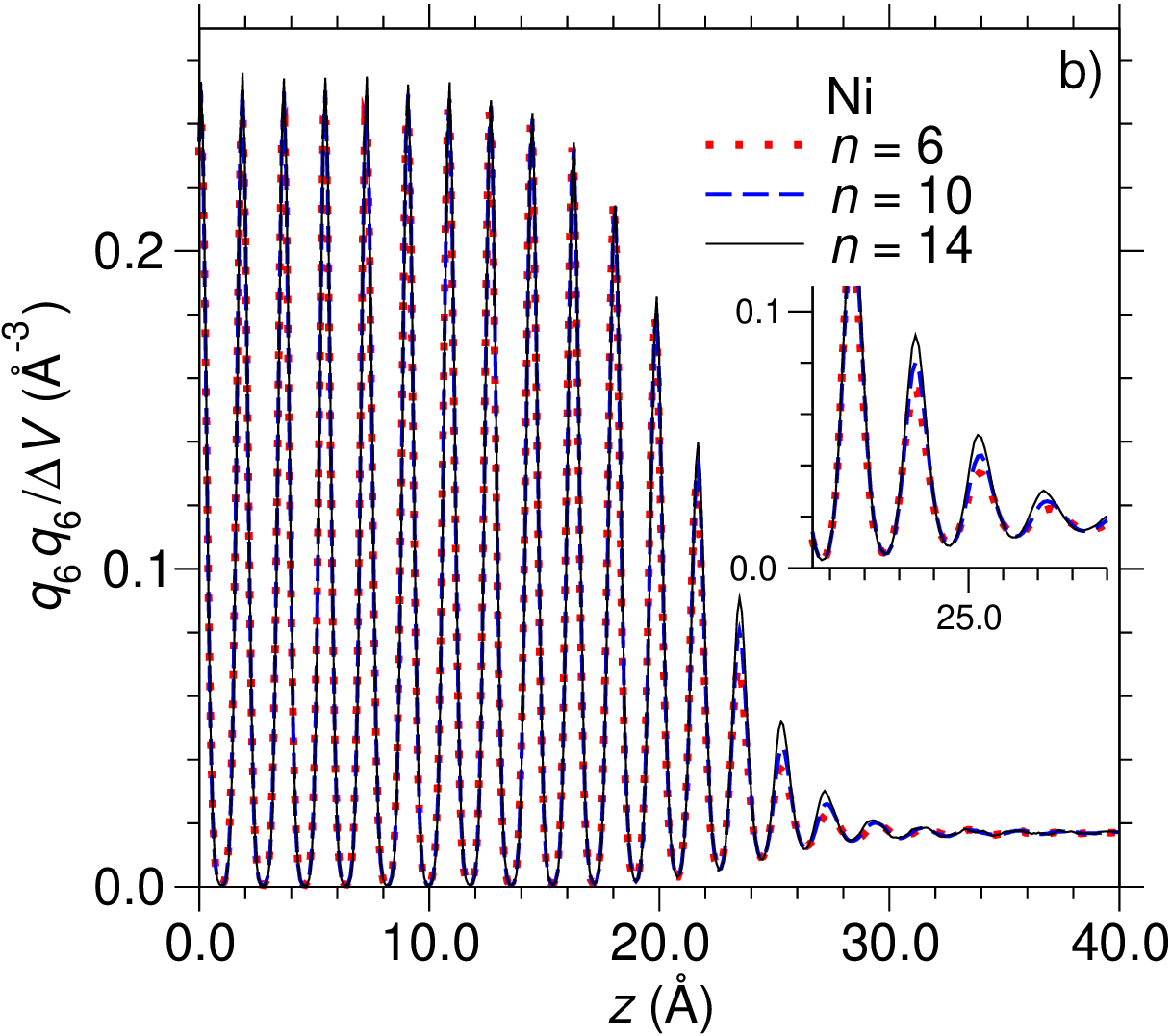}
\caption{Profile for the order parameter $q_6q_6$, a) for hard spheres
and b) for Ni. The inserts provide a magnification of the interface 
region.}
\label{fig4}
\end{center}
\end{figure}
\begin{figure}
\begin{center}
\includegraphics[width=0.45\textwidth,clip]{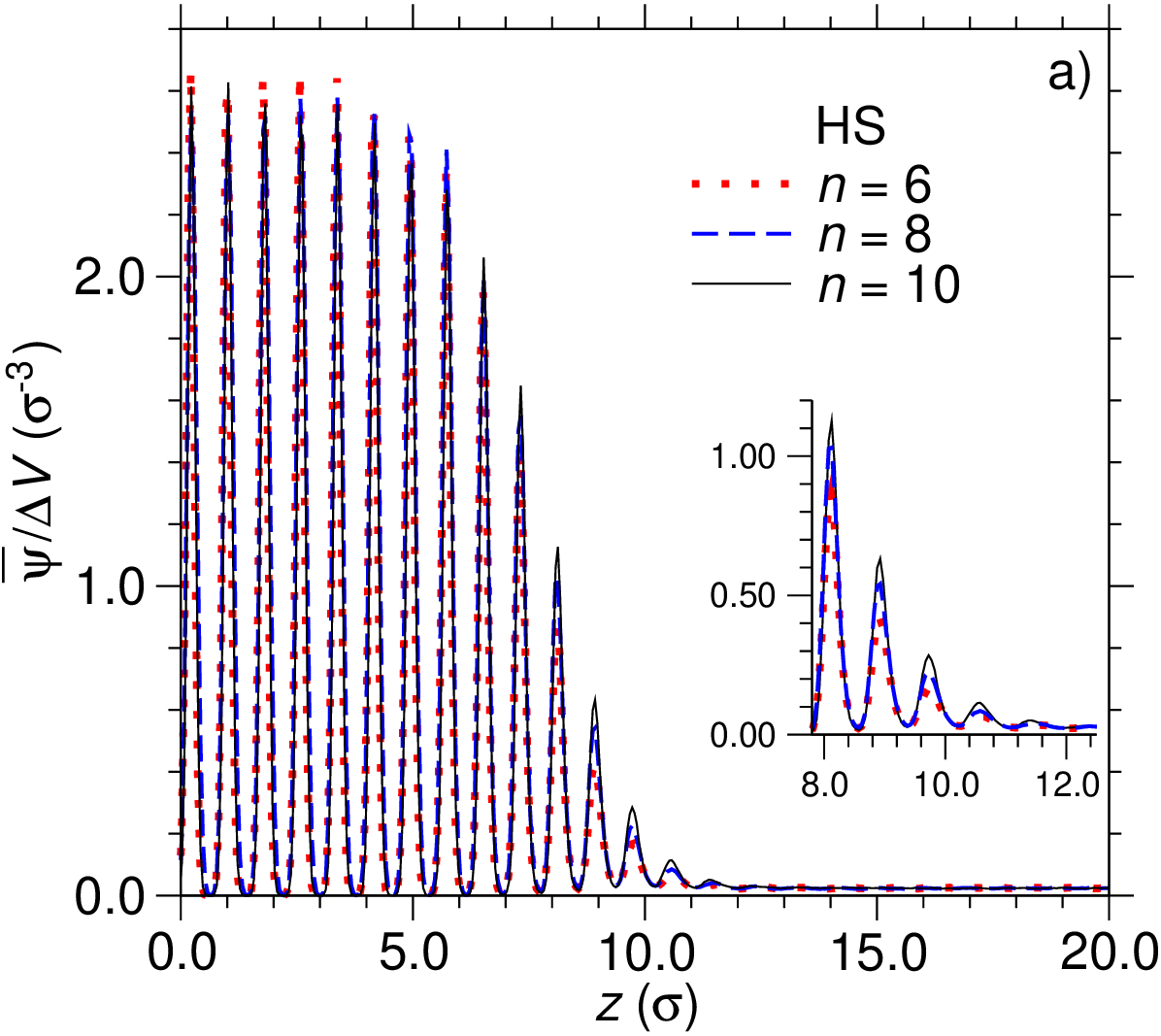}
\includegraphics[width=0.45\textwidth,clip]{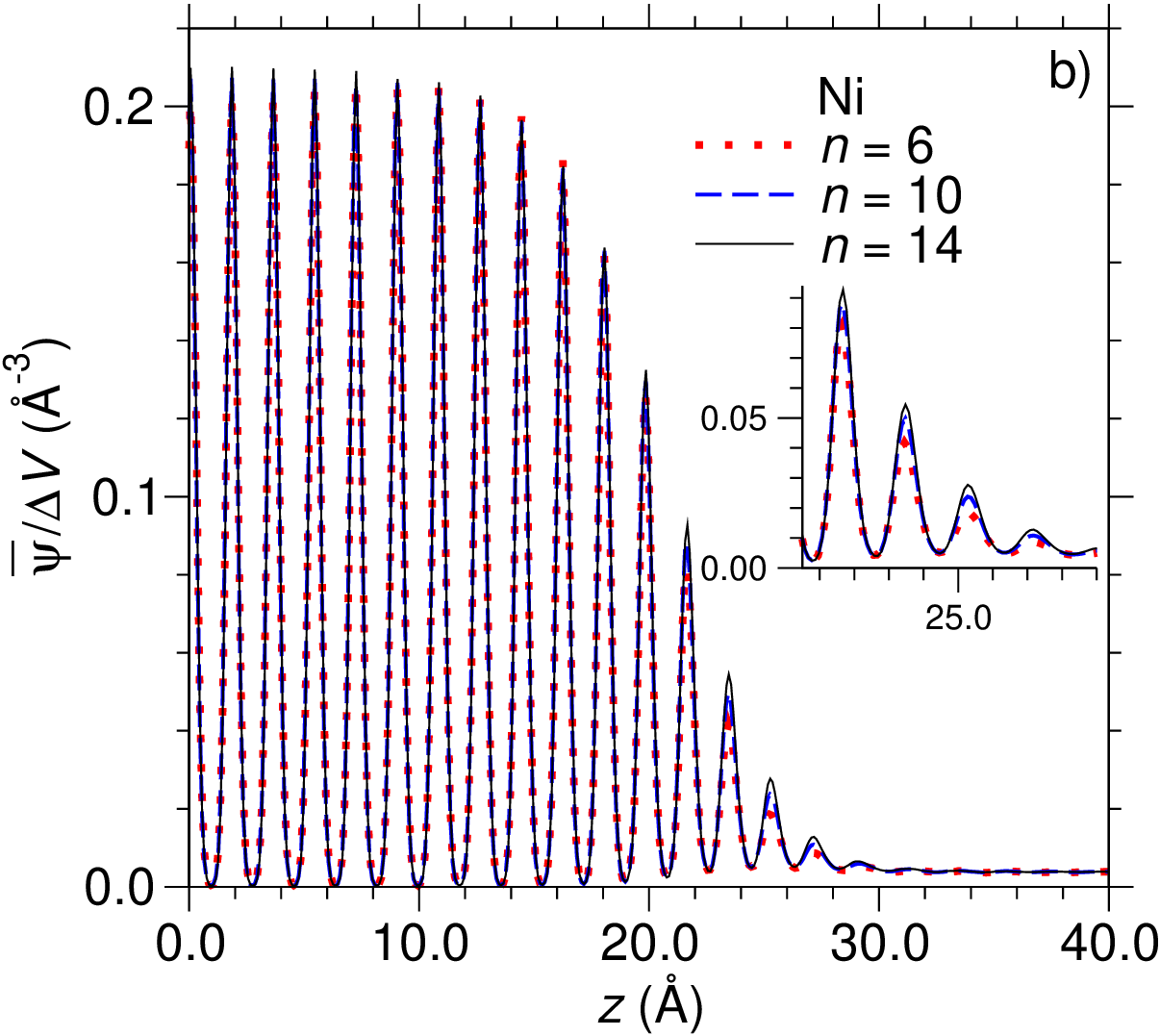}
\caption{Profile for the order parameter $\bar{\Psi}_r$, a) for hard spheres
and b) for Ni. The inserts provide a magnification of the interface region.}
\label{fig5}
\end{center}
\end{figure}
Figure \ref{fig3} displays local order parameter distributions for the
pure liquid and the pure solid phases. The distributions indicate that
the considered order parameters are well-suited to distinguish liquid
from solid particles. For the calculation of the local order parameters,
we used time-averaged particle positions. To this end, for the hard
sphere system, positions were averaged over 50 MC cycles. For Ni,
the phonon's degrees of freedom lead to a significant shift of the
order parameter distributions.  Using a time interval of 0.1\,ps for
the averaging of the particle positions in Ni, compared to an ideal
fcc crystal the order parameter distributions for the crystal phase
are broader and tend to shift to smaller values of the order parameter
(see Fig.~\ref{fig3}). For a time average over 1\,ps the order parameter
distributions are very similar to those for the hard spheres. However,
since in Ni at $T_{\rm m}$ a time scale of 1\,ps is already close to
the time scale of particle diffusion in the melt, we have used a time
averaging over 0.1\,ps for the analysis of the interfacial properties
that are presented in the following.

\begin{figure}
\begin{center}
\includegraphics[width=0.45\textwidth,clip]{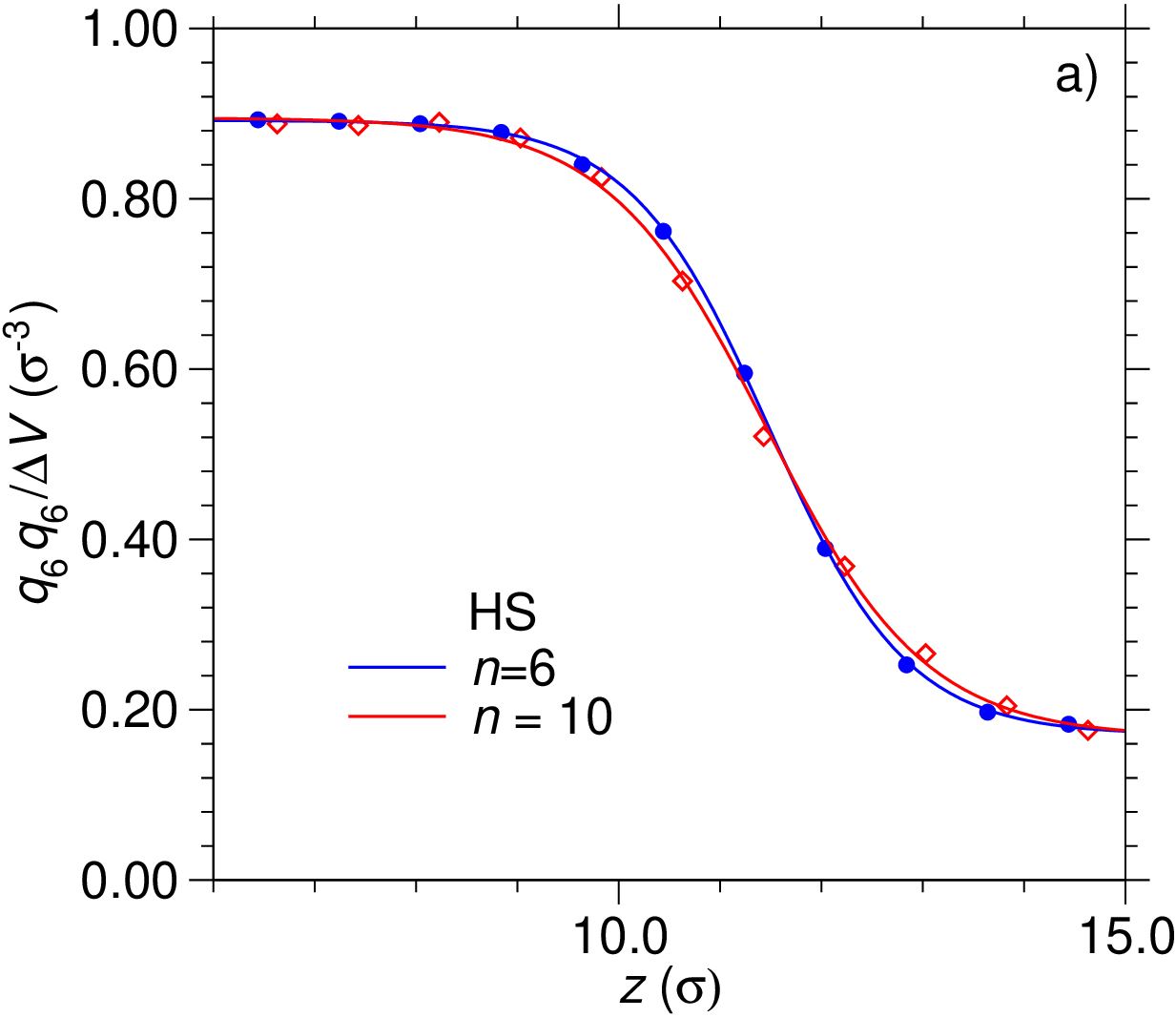}
\includegraphics[width=0.45\textwidth,clip]{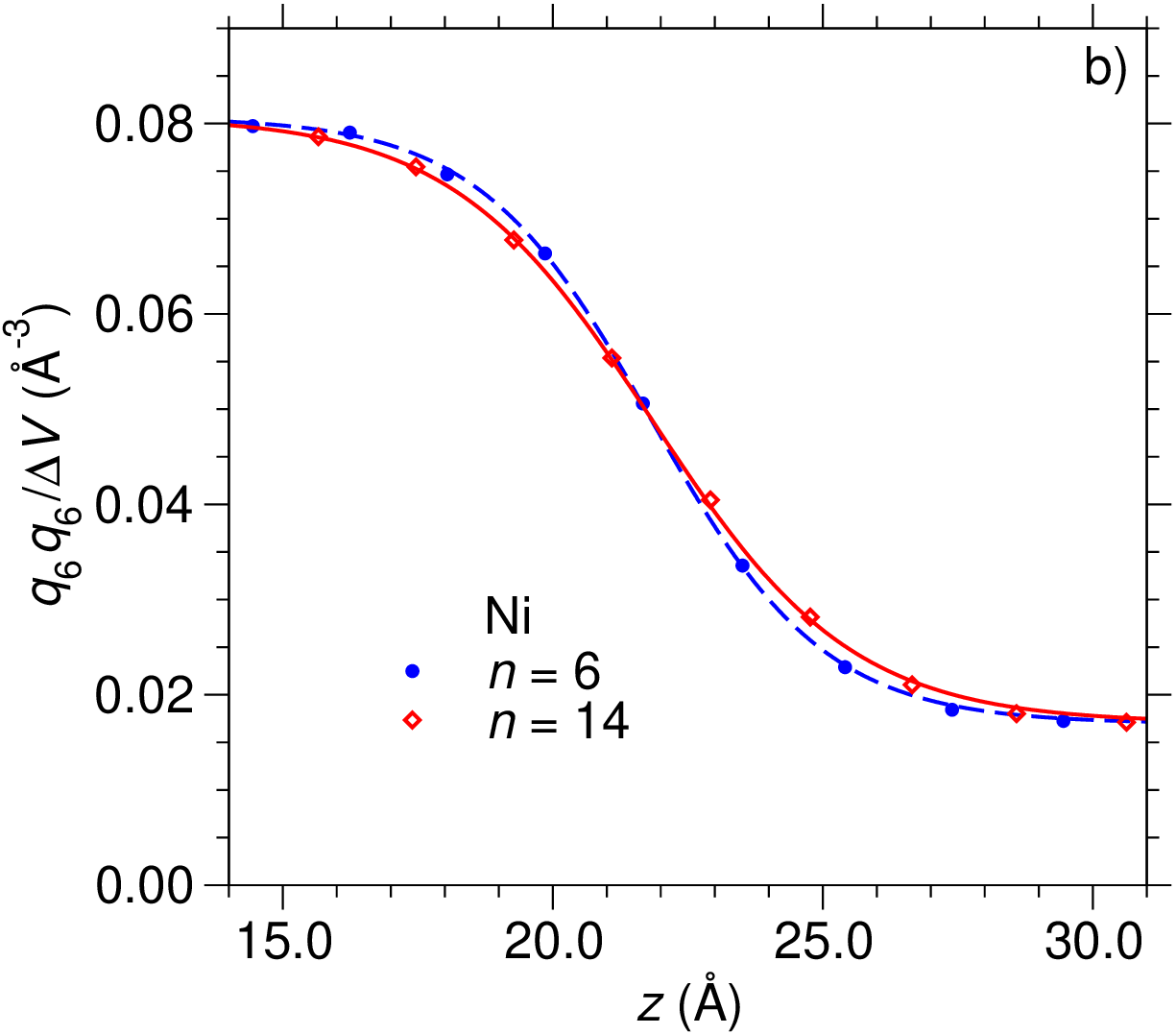}
\caption{Coarse grained order parameter profile for $q_6 q_6$ at the indicated
system sizes, a) for hard spheres and b) for Ni. The solid lines are fits to 
a hyperbolic tangent function (Eq.\ref{eq_mf}), see text.}
\label{fig6}
\end{center}
\end{figure}
In Figs.~\ref{fig4} and \ref{fig5}, profiles of the local order parameters
are shown, i.e.~the sum of the order parameter of the particles contained
within a slice transversal to the solid-liquid interface divided by the
volume of the slice $\Delta V=L^2\Delta z$. The profiles are calculated
with a resolution (bin size) of 0.02\,$\sigma$ for the hard spheres and
0.1\,\AA~for Ni.  The order parameter profiles show similar features as
the density profiles.  However, finite-size effects seem to be revealed
in a more pronounced manner.  Clearly, the height of the peaks in the
interface region slightly increases with increasing system size.

\begin{figure}
\begin{center}
\includegraphics[width=0.45\textwidth,clip]{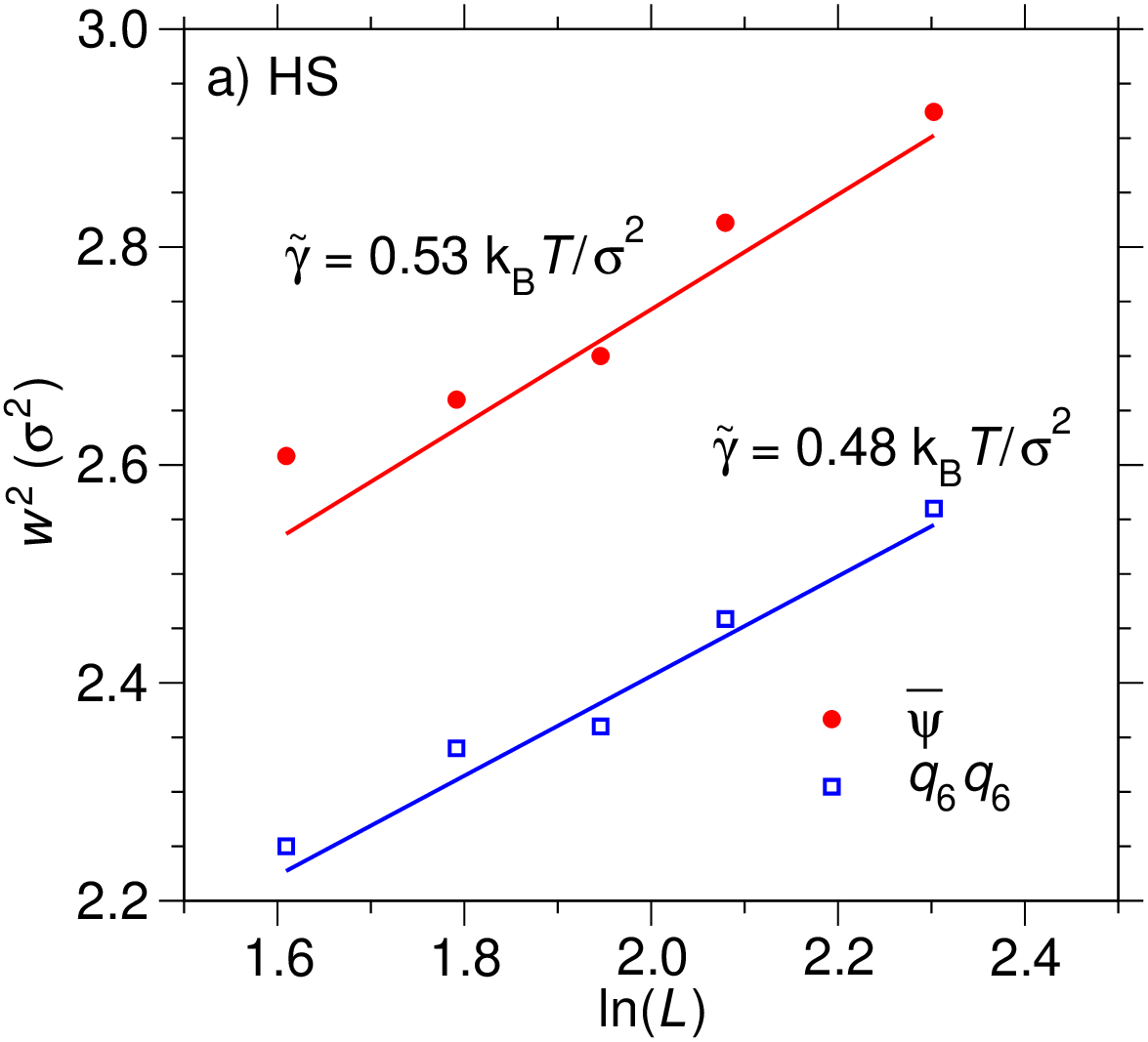}
\includegraphics[width=0.45\textwidth,clip]{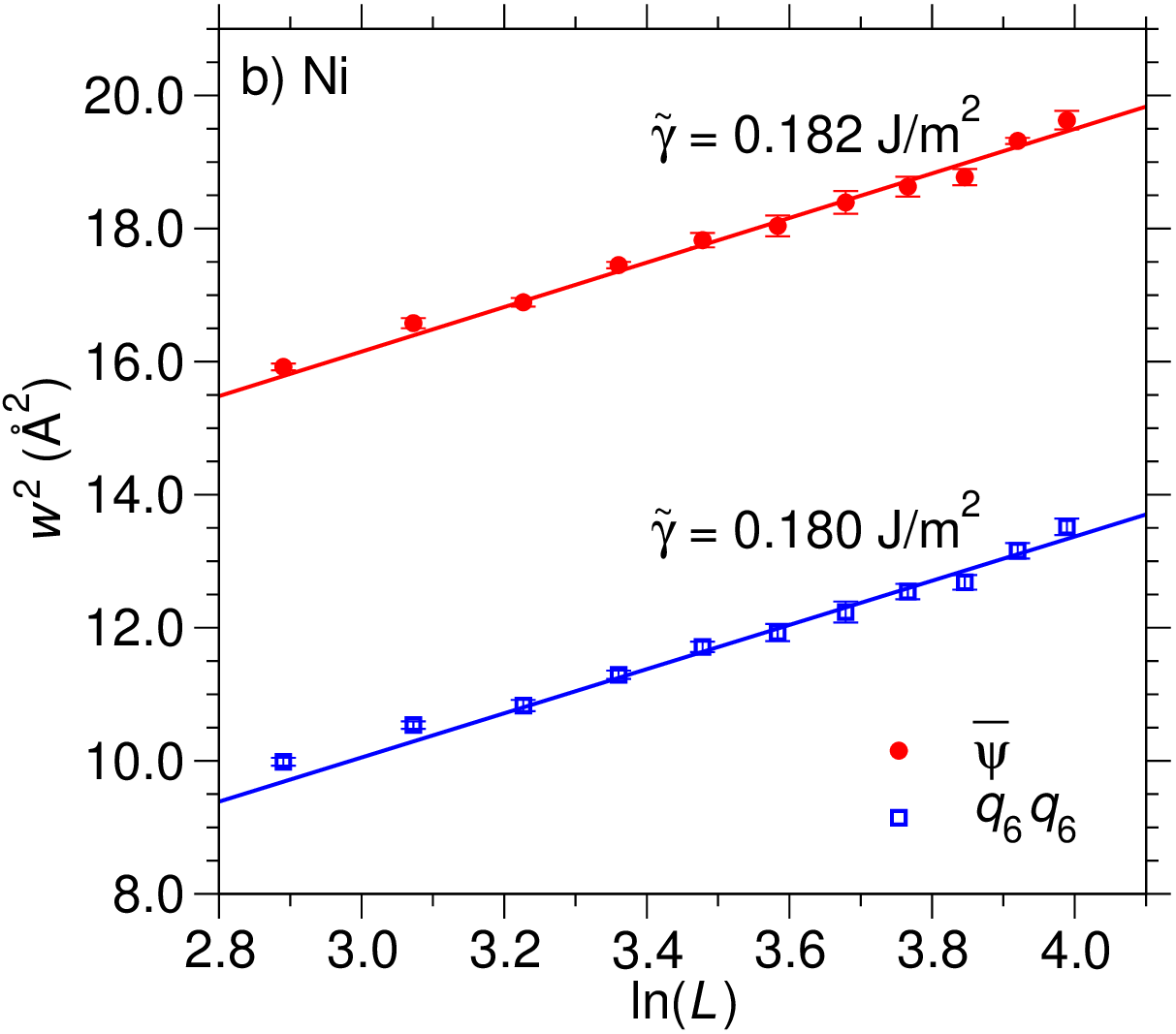}
\caption{Mean-squared width $w^2$ as a function of $\ln L$, a) for hard spheres 
and b) for Ni. The values for $\tilde{\gamma}$ are obtained from the fits
(solid lines).}
\label{fig7}
\end{center}
\end{figure}
To compute coarse-grained profiles we identify the minima in the fine-grid
profiles of Figs.~\ref{fig4} and \ref{fig5}. These minima define the
borders of nonuniform bins that match the crystalline layers. Then,
we compute the average value of the order parameter in each of the
latter bins. Examples for the resulting coarse-grained order parameter
profiles for the case of $q_6 q_6$ are displayed in Fig.~\ref{fig6}.
Here, the solid lines are fits with a hyperbolic tangent function,
$\phi(z) = A - B \tanh[(z-z_0)/w]$ (where $A$ and $B$ are parameters
related to the bulk values of the order parameter, and $z_0$ and $w$
are the interface position and its effective width, respectively). Both
for the hard spheres and Ni, the latter fits indicate that the width $w$
is larger for the big systems, as expected from CWT.

\subsection{Estimate of the interfacial stiffness}
In Fig.~\ref{fig7}, the mean-squared width $w^2$, as obtained from the
fits to the coarse-grained profiles for $q_6 q_6$ and $\bar{\Psi}$, is plotted
as a function of $\ln L$. The plot confirms the logarithmic increase of
$w^2$ with the lateral system size, as predicted by CWT.

From the fits with Eq.~(\ref{eq_width}), we estimate for the (100)
orientation $\tilde{\gamma}=0.50 \pm 0.05 \, k_B T/ \sigma^{2}$ for
the hard-sphere system and  $\tilde{\gamma}=0.18 \pm 0.01$\,J/m$^{2}$
for Ni.  These values roughly agree with previous estimates, obtained by
other methods.  For hard spheres, Davidchack and Laird~\cite{davidchack00}
found $\tilde \gamma =0.57 \, k_BT/\sigma^2$ using a thermodynamic
integration approach, while Mu et al.~\cite{mu05} obtained $\tilde \gamma
\simeq 0.62 \, k_BT/\sigma^2$ from the analysis of the capillary wave
spectrum. However, Davidchack et al.~\cite{laird06} later criticized their
result as being biased and rather suggested $\tilde \gamma = 0.56 \pm 0.02
\, k_BT/\sigma^2$ when averaged over all interface orientations. For the
(100) orientation, they suggest $\tilde \gamma\simeq 0.44 \pm 0.03 \,
k_BT/\sigma^2$.  However, their actual data reveal huge fluctuations,
and the judgement of the actual accuracy may need reanalysis.  For Ni,
Hoyt et al.~\cite{hoyt01} determined the interfacial stiffness for
the (100) orientation (and other orientations of the Ni fcc phase).
Using a different EAM model and the analysis of the capillary wave
spectrum to measure $\tilde{\gamma}$, they obtained $\tilde{\gamma}
\approx 0.23$\,J/m$^2$, which is slightly larger than our result.

\section{Summary and Outlook}
In this paper, we have presented a comparative study of melt-crystal
interfaces for hard spheres and an embedded atom model for nickel. These
rather diverse systems have been studied in analogous geometries, namely
$L\times L \times L_z$ rectangular simulation boxes with periodic boundary
conditions, at conditions where a crystalline slab, separated by two $L
\times L$ interfaces oriented perpendicular to the $z$-direction, coexists
with the fluid phase.  The motivation for this study was to provide a
better understanding of the information that one can extract from the
simulation study of such interfaces, paying particular attention to finite
size effects, and to limitations of the accuracy which are inherently due
to the simulation setup.  In fact, due to the translational invariance of
the simulation geometry as a whole, the center of mass of the crystalline
part is not fixed in space, but may fluctuate and diffuse along the
$z$-axis. In addition, at phase coexistence in a finite box, fluctuations
may occur when the size of the crystal (volume fraction of the box that
is crystallized) changes. As a consequence, in each configuration that is
analyzed one must locate the precise position along the $z$-axis where
the lattice planes are (this was done via a fine-grid coarse graining)
and then the time averages are found in such a way that the positions
of lattice planes and interface centers coincide. It is clear that this
is a delicate procedure and hence there is the need to watch out for
possible systematic errors, which are not necessarily equally important
for different kinds of systems, and for different simulation methods
(e.g. MC and MD). In view of these caveats, it is gratifying to state
that with the methods described in this paper, these problems seem
reasonably well under control. Indeed, we find that the main difficulty
in the interpretation of our results for the interfacial profiles and
their width is the broadening by the capillary waves. This phenomenon,
though well-known for vapor-liquid interfaces, has found comparatively
little attention for the melt-crystal interface. Our results imply that
the capillary wave broadening (for rough, non-facetted crystal surfaces)
is present and important: while it makes a naive direct comparison with
DFT calculations of interfacial profiles obsolete, it yields a relatively
straightforward method for extracting the interfacial stiffness and the
accuracy of this method seems to be competitive to other approaches.

As a next step we plan a detailed comparison with the alternative
method where the Fourier spectrum of interfacial fluctuations is
analyzed. For vapor-liquid type systems, such comparisons can be found
in the literature, but for the melt-crystal interface a comprehensive
comparative assessment of different methods still is lacking. Of course,
for applications in crystal nucleation and growth phenomena very accurate
estimates for the interfacial stiffness are indispensable.

{\bf Acknowledgments:} We thank Lorenz Ratke for a critical reading
of the manuscript. We are grateful to the German Science
Foundation (DFG) for financial support in the framework of the SPP 1296.
We acknowledge a substantial grant of computer time at the J\"ulich
multiprocessor system (JUMP) of the John von Neumann Institute for
Computing (NIC) and at the ZDV of the University of Mainz.

\section*{References}

\end{document}